\documentclass[aps,floatfix,showpacs,nofootinbib,amsmath,amssymb]{revtex4}
\usepackage[dvips]{graphicx}
\usepackage{color}
\usepackage{verbatim} 

\begin{document}


\title{Description of our cosmological spacetime as a perturbed
conformal Newtonian metric and implications for the backreaction proposal for
the accelerating universe}

\author{Edward W.\ Kolb} \email{rocky.kolb@uchicago.edu} 
\affiliation{ Department of Astronomy and Astrophysics, Enrico Fermi
Institute, and  Kavli Institute for Cosmological Physics, the University of
Chicago, Chicago, IL 60637-1433, USA}

\author{Valerio Marra} \email{valerio.marra@pd.infn.it}
\affiliation{Dipartimento di Fisica ``G.\ Galilei'' Universit\`{a} di Padova,
INFN Sezione di Padova, via Marzolo 8, Padova I-35131, Italy}
\affiliation{Kavli Institute for Cosmological Physics, the University of 
Chicago, Chicago, IL 60637-1433, USA}

\author{Sabino Matarrese} \email{sabino.matarrese@pd.infn.it}
\affiliation{Dipartimento di Fisica ``G.\ Galilei'' Universit\`{a} di Padova,
INFN Sezione di Padova, via Marzolo 8, Padova I-35131, Italy}

\begin{abstract}

It has been argued that the spacetime of our universe can be accurately
described by a perturbed conformal Newtonian metric, and hence even large
density inhomogeneities in a dust universe can not change the observables
predicted by the homogeneous dust model.  In this paper we study a spherically
symmetric dust model and illustrate conditions under which large spatial
variations in the expansion rate can invalidate the argument.

\end{abstract}


\maketitle

\section{Introduction}

In cosmology, one models the evolution and observables associated with an
inhomogeneous universe of density $\rho(\vec{x})$ and expansion rate
$H(\vec{x})$ by employing a Friedmann-Lema\^{\i}tre-Robertson-Walker (FLRW)
homogeneous/isotropic model of density $\rho = \langle\rho(\vec{x})\rangle$,
where $\langle\cdots\rangle$ denotes some suitably defined spatial average. 
One then assumes that the expansion rate and cosmological observables are those
obtained in the corresponding FLRW model.

One of the proposals to explain ``dark energy'' calls into question this
long-standing (86-year) procedure.  The idea is that the expansion rate and
cosmological observables of a suitably inhomogeneous universe containing only
dust, if analyzed within the framework of a homogeneous model, seems to behave
as if the stress tensor also contains a fictitious negative-pressure fluid
(\textit{i.e.,} dark energy).

Although this proposal is conservative in the sense that it does not involve a
cosmological constant of incredibly small magnitude (presumably rationalized on
some anthropic basis), a scalar field of unbelievably small mass, or an
entirely unmotivated modification of general relativity, it is rather
revolutionary because it implies that there is no dark energy and the expansion
of the universe does not actually accelerate (at least, not in the usual
sense).

At present, the idea that the backreaction\footnote{By the backreaction idea we
mean both that inhomogeneities influence cosmological observables (weak
backreaction) and that inhomogeneities influence the average/background scale
factor (strong backreaction). Weak and strong backreactions are, however,
likely connected if we assume that averages should be performed along the
light cone. } of inhomogeneities accounts for the observational evidence
usually attributed to dark energy is more of a concept than a predictive
model.  However, it is generally agreed that if the proposal is to be relevant,
nonlinearities are required.  

There have been many criticisms of this approach.  One of them
\cite{Ishibashi:2005sj,Gruzinov:2006nk} is based on the claim that even in the
presence of highly nonlinear density perturbations ($\delta\rho/\rho\gg1$) the
metric for our universe can everywhere be written  as a perturbed conformal
Newtonian metric of the form\footnote{We will refer to a metric written in the
form of Eq.\ (\ref{newtpert}), satisfying the stated conditions, as the
\textit{perturbed conformal Newtonian} metric.}  
\begin{equation}
ds^2=a^2(\tau)\left[-(1+2\psi)d\tau^2 + (1-2\psi)\gamma_{ij} dx^idx^j \right]  ,
\label{newtpert}
\end{equation}
where $d\tau = dt/a$ is conformal time, $\gamma_{ij}$ is a metric of a
three-space of constant curvature, and  $\psi$ satisfies the Newtonian
conditions $|\psi| \ll 1$,  $\left| \partial\psi/\partial t \right|^2 \ll 
a^{-2}D^i\psi D_i\psi$, and $\left(D^i\psi D_j\psi\right)^2\ll
\left(D^iD^j\psi\right)D_iD_j\psi$.  The covariant derivative with the metric
$\gamma_{ij}$ is denoted by $D_i$.  The usual statement is that in the dust
case one is allowed to use the perturbed conformal Newtonian metric either in
the linear regime (\textit{i.e.,} perturbations of every quantity being small)
or in the weak-field (Newtonian) regime.\footnote{But already at second order
in perturbation theory the spatial part of the metric differs from the temporal
part of the metric simply from terms quadratic in the peculiar velocities
(\textit{i.e.,} an effective anisotropic stress term is in the stress-energy
tensor).} 

The claim is that if the metric can be written in the above form and satisfies
the above conditions, even in the presence of large inhomogeneities, any
cosmological observable will be the same as  the cosmological observable
calculated with $\psi=0$, \textit{i.e.,} in the homogeneous/isotropic model.
This has been described as a ``no-go'' theorem that backreactions can not
\textit{in principle} account for the observations.  

While it is impossible to consider the most general inhomogeneous
solutions, there are spherically symmetric inhomogeneous dust solutions, which
are not perturbations of Einstein-de Sitter, that can be constructed to give
observables similar to $\Lambda$CDM models.  These models serve as a
counterexample to the no-go argument.  In this paper we will show why these
models can not be described in terms of a conformal Newtonian metric perturbed
about a spatially flat background, and attempt to understand the implications
for the backreaction proposal.  

Indeed, while it may turn out that backreactions are not the answer, we argue that
assuming the results of measurements of the luminosity distance as a function
of redshift usually interpreted as an accelerated expansion, the metric
describing our universe can not be written in the form of a perturbed conformal
Newtonian metric where $a(t)$ is calculated from the homogeneous dust model. 
In other words, if the expansion history of the universe is well described by
the $\Lambda$CDM model, then perturbing about an Einstein-de Sitter model by
the perturbed conformal Newtonian metric of Eq.\ (\ref{newtpert}) is
inappropriate, because Einstein-de Sitter would be the wrong background. This
is because of large peculiar velocities with respect to the background
Einstein-de Sitter space.  So if inhomogeneities are responsible for the
observables usually attributed to dark energy, the universe can not be obtained
by small perturbations of the Einstein-de Sttter model.  In other words, the
reason we interpret the observations as evidence for dark energy and
acceleration of the universe is that we are comparing the observables to
observables computed in the wrong background.

As we will discuss, the reason is that the proper meaning of ``peculiar'' is
``after subtraction of a background Hubble flow term.'' We will argue that
large peculiar velocities must be present if the backreaction program works,
and the peculiar velocities are not related to ``local'' departures from the
Hubble flow that would show up as large velocity dispersions.

As an explicit example, consider the argument of Ref.\ \cite{Kolb:2005da}. They
propose that the backreaction of inhomogeneities in a dust universe modifies
the evolution of the effective volume-averaged scale factor and results in an
evolution of the volume expansion that resembles a $\Lambda$CDM model, rather
than the unperturbed spatially flat dust model. If one would write the metric
for such a perturbed universe in terms of a perturbed conformal Newtonian
metric, then one would have to use $a(t)$ described by a $\Lambda$CDM model,
\textit{not} the $a(t)$ from an unperturbed spatially flat dust model.  If one
would attempt to express the metric in terms of a perturbed metric with $a(t)$
described by a spatially flat dust model, then there would be enormous peculiar
velocities in the Hubble flow.

We explore the issue by considering the spacetime of a particular
Lema\^{\i}tre-Tolman-Bondi spherically symmetric model. In the model we study,
the observer is centered in a large (Gpc) underdense region in an
asymptotically Einstein-de Sitter universe.  This model has large, but not
highly nonlinear, density and curvature inhomogeneities.  We show by explicit
calculation that the metric \textit{can not} be expressed as a perturbed
conformal Newtonian metric during the entire evolution.  We also show that the
reason for this can be traced to large peculiar velocities that develop with
respect to the Einstein-de Sitter background.  

In the next Section we describe the relevant features of the Einstein-de
Sitter  and Lema\^{\i}tre-Tolman-Bondi models.  In Section \ref{LTBPG} we
develop the formalism to express the original Lema\^{\i}tre-Tolman-Bondi metric
in the form of a perturbed conformal Poisson metric. In Section
\ref{Discussion} we discuss the reason for the numerical results of the
previous Section.  Finally, in Section \ref{Conclusions} we present our
conclusions.

\section{Einstein--de Sitter (EdS) and Lema\^{\i}tre--Tolman--Bondi (LTB) 
models \label{ltbeds}}

In this Section we will establish the notation for the homogeneous, isotropic,
and spatially-flat cosmological model (the Einstein--de Sitter model). We will
then briefly introduce the formalism of LTB spherically symmetric cosmological
models in the synchronous gauge, and turn to a particular LTB model with values
of cosmological observables similar to the $\Lambda$CDM model. 

\subsection{The EdS model}

The EdS model describes  a spatially-flat, matter-dominated universe.   In
order to connect with the LTB solution of the next section, we can express the
line element in terms of the scale factor $a(t)$: 
\begin{equation} 
ds^2=-dt^2+a^2(t) dr^2 + a^2(t)r^2 \, d\Omega^2 . 
\end{equation}

The Friedmann equation and its solutions are ($a_0$ is the present scale 
factor--in general, the subscript $0$ will denote the present value of a
quantity and $\kappa=8\pi G$)
\begin{eqnarray}
H^2(t) & = & \left(\dot{a}/a\right)^2=\kappa\rho(t)/3=H_0^2(a_0/a)^3 
\nonumber \\
a(t) & = & a_0\left(t/t_0\right)^{2/3}\nonumber \\
\rho(t) & = &\rho_0(a_0/a)^{3}  .
\end{eqnarray}

At any epoch, the age of the universe is $t=2H^{-1}/3$.  We also note that the
present distance to the horizon is $d_H=2H_0^{-1}$.

\subsection{The LTB model \label{theltb}}

The LTB model \cite{Lemaitre:1933gd,Tolman:1934za,Bondi:1947av} is based on the
assumptions that the system is spherically symmetric with purely radial motion
and the motion is geodesic without shell crossing (otherwise we could not
neglect the pressure). The line element in the synchronous gauge can be written
in the form
\begin{equation}
\label{ltbmetric}
ds^2=-dt^2 + \frac{R'^2(r,t)}{W^2(r)} dr^2 + R^2(r,t) \, d\Omega^2 .
\end{equation}
Here, the prime superscript denotes $d/dr$.  We will denote $d/dt$ by an
overdot.  The function $W(r)$ is an arbitrary function of $r$.  We will also
find it useful to define equivalent functions $\beta(r)$ and $k(r)$, related 
to $W(r)$ by
\begin{equation}
W^2(r)=1+\beta(r) = 1-k(r)r^2.
\end{equation}
One can recover the usual Friedmann-Robertson-Walker metric by the replacements
\begin{eqnarray}
\label{recF}
R(r,t) & \rightarrow & r a(t) \nonumber \\
k(r) & \rightarrow & \pm1, \textrm{ or } 0.
\end{eqnarray}

In spherically symmetric models, in general there are two expansion rates: an
angular expansion rate, $H_\perp\equiv \dot{R}(r,t)/R(r,t)$, and a radial
expansion rate, $H_r\equiv \dot{R}'(r,t)/R'(r,t)$. 

Assuming a dust equation of state, the Einstein equations may be expressed as
\begin{eqnarray}
H_\perp^2+2H_rH_\perp -\frac{\beta}{R^2}-\frac{\beta'}{RR'} & = & \kappa\rho
\nonumber \\
6\frac{\ddot{R}}{R} + 2H_\perp^2-2\frac{\beta}{R^2} - 2H_rH_\perp
+ \frac{\beta'}{RR'} & = & -\kappa\rho .
\end{eqnarray}
These are the generalization of the Friedmann equations for a
homogeneous/isotropic universe to a spherically symmetric inhomogeneous
universe.

Adding the equations results in an equation that can be easily integrated to
yield the dynamical equation for $R(r,t)$:
\begin{equation}
\label{rdot}
\dot{R}(r,t)=\sqrt{\beta(r)+\frac{\alpha(r)}{R(r,t)}} ,
\end{equation}
where $\alpha(r)$ is a function of $r$ related to the density $\rho(r,t)$ by
\begin{equation}
\kappa\rho(r,t)=\frac{\alpha'(r)}{R^2(r,t)R'(r,t)} .
\end{equation}
Equation (\ref{rdot}) can be differentiated to yield the dynamical equation 
for $R'(r,t)$:
\begin{equation}
\label{rpdot}
\dot{R}'(r,t) = \frac{\beta'(r)+\alpha'(r)/R(r,t)-\alpha R'(r,t)/R^2(r,t)}
{2\dot{R}(r,t)} .
\end{equation}

Since we will eventually express the LTB solution using a perturbed conformal 
Poisson metric, it is convenient to use instead of time a new independent
variable related to the EdS scale factor: $x=a/a_0$.  We also introduce new
dependent variables $Y=R/ar$ and $Z=R'/a$.  Therefore, our variables are
\begin{eqnarray}
x & = & \frac{a}{a_0} \nonumber \\
Y(x,r) & = & \frac{R}{ar} \nonumber \\
Z(x,r) & = & \frac{R'}{a} .
\end{eqnarray}
The solution $Y=Z=1$ corresponds to the EdS model.  The dynamical equations for
$Y$ and $Z$ are
\begin{eqnarray}
\frac{dY}{dx} & = & \frac{Y}{x}\left[
\sqrt{\frac{B_1x}{Y^2}+\frac{A_1}{Y^3}}-1 \right] \nonumber \\
\frac{dZ}{dx} & = & \frac{Z}{x}\left[ \frac{1}{2Z} 
\frac{xB_2+A_2/Y-A_1Z/Y^2}{\sqrt{B_1x+A_1/Y}} -1\right], \label{dynamicaleqs}
\end{eqnarray}
where the functions $A_1(r)$, $A_2(r)$, $B_1(r)$, and $B_2(r)$ are related to
$\alpha(r)$, $\beta(r)$ and their derivatives by
\begin{eqnarray}
\alpha(r)  = H_0^2a_0^3r^3A_1(r) \nonumber \\
\alpha'(r) = H_0^2a_0^3r^2A_2(r) \nonumber \\
\beta(r)   = H_0^2a_0^2r^2B_1(r) \nonumber \\
\beta'(r)  = H_0^2a_0^2r  B_2(r).
\end{eqnarray}
Here, $H_0=H_\perp(t_0, r\rightarrow\infty) = H_r(t_0, r\rightarrow\infty) =
100h \textrm{km s}^{-1}\textrm{Mpc}^{-1}$ is the present EdS expansion rate. 

Since each shell labeled by $r$ evolves independently, we can choose initial
conditions to be specified at different times for different shells.  However,
for simplicity we choose to specify initial conditions at the same time for all
shells. We will choose the initial time to be recombination for the EdS model,
$x^{-1}=1100$.  The initial value of $R$ (or $Y$) can be chosen freely with a
suitable rescaling of $r$.  We choose $Y=Z=1$ at $x^{-1}=1100$ to match to the
EdS solution as $r\rightarrow\infty$.

A particular LTB model is generated by a choice for $\alpha(r)$ and $\beta(r)$.
For a discussion of the free functions in LTB models, see Ref.\
\cite{Marra:2007pm}. For other works based on LTB solutions aiming at showing
the effect of inhomogeneities, see
\cite{Mustapha:1998jb,Celerier:1999hp,Iguchi:2001sq,Apostolopoulos:2006eg,Vanderveld:2006rb,Chung:2006xh,
Enqvist:2006cg,Biswas:2006ub,Marra:2007pm,Marra:2007gc,Marra:2008xi,
Marra:2008sy,Brouzakis:2007zi,Biswas:2007gi,Brouzakis:2008uw}.
Our goal here is to study a LTB model with observables similar to the standard
$\Lambda$CDM model.  We will adopt the particular model of Alnes, Amarzguioui,
and Gr{\o}n (AAG) \cite{Alnes:2005rw,Alnes:2006uk,GarciaBellido:2008nz}, but
with slightly different parameters.  The model is EdS at large $r$, the
observer is in the center of an underdense (compared to the EdS density)
region, and a there is region of density larger than the EdS value to
compensate the underdense region.  

\begin{figure}
\begin{center}
\includegraphics[width=16cm]{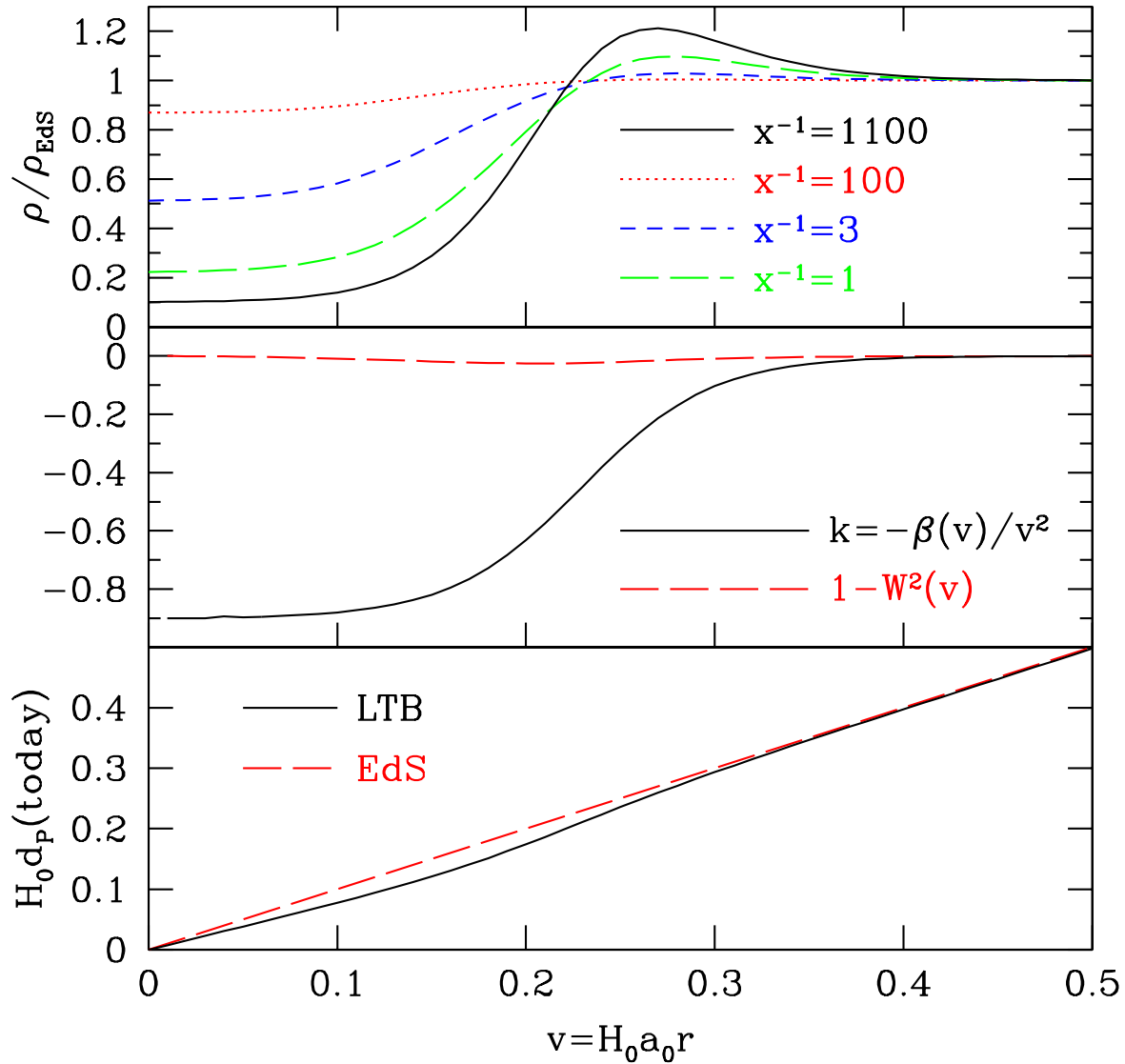}
\caption{Upper figure: The density compared to the density of the EdS model as
a function of $v=H_0a_0r$ for various values of $x=a/a_0$.  Middle figure: The
functions $k=-\beta(v)/v^2$ and $-\beta(v) = 1-W^2(v)$ as a function of
$v=H_0a_0r$. Notice that $W^2$ is always very close to unity (an approximation
we will employ later). Bottom figure: The current proper distance as a function
of the  comoving coordinate $v=H_0a_0r$ for the LTB model and the EdS model 
($H_0^{-1}=5878\textrm{Mpc}$).\label{rbd}}
\end{center}
\end{figure}

In the AAG model $\alpha(r)$ and $\beta(r)$ are given by
\begin{eqnarray}
\alpha(r) & = &
H_0^2a_0^3r^3\left[1-\frac{\Delta\alpha}{2}
\left(1-\tanh\frac{r-r_0}{2\Delta r}\right)\right] \nonumber \\
\beta(r) & = & H_0^2a_0^2r^2 \frac{\Delta\alpha}{2}
\left(1-\tanh\frac{r-r_0}{2\Delta r}\right) .
\end{eqnarray}
Alnes, Amarzguioui, and Gr{\o}n choose $h=0.51$, $\Delta\alpha=0.9$, 
$a_0r_0=1.35$ Gpc, $\Delta r = 0.4r_0$.  With that choice of parameters, at the
origin at present $h_\perp=h_r=0.65$. They found that this choice of parameters
reproduces the luminosity distance--redshift diagram expected for a
$\Lambda$CDM model.  That choice of parameters results in a ``weak''
singularity at the origin \cite{Vanderveld:2006rb}, as can be see by the fact
that $\rho'(r=0)\neq0$.\footnote{What Ref.\ \cite{Vanderveld:2006rb} actually
claims is that if there is no weak singularity at the center, then for an
observer at the center, $d\Delta(m-M)/dz$ cannot be positive at $z=0$. 
However, it is possible to have $d\Delta(m-M)/dz>0$ for off-center observers.} 
While this is not a serious issue (the observer could be slightly displaced
from the origin or a slightly different density profile could be used) we will
employ different parameters than AAG to mitigate the weak singularity.  We do
this simply by the choice $\Delta r=0.15r_0$.  This results in the present
expansion rate at the origin of $h_\perp=h_r=0.67$.

The density compared to the EdS density is shown in Fig.\ \ref{rbd}.  Notice
that at the origin $\rho'$ vanishes. The curvature function $k(r)$ is also
shown in Fig.\ \ref{rbd}.  The fact that as $r$ approaches $0$, $k$ is negative
and $\rho$ is much smaller than the EdS value, implies that the evolution deep
in the void will resemble a negatively-curved, empty universe.  For
sufficiently large $r$, $k$ approaches $0$ and $\rho$ approaches the EdS value,
so far from the void the universe evolves as an EdS model universe.\footnote{We
will refer to the value of $r$ or $v$ when the solution is very near the EdS
solution as $r_\textrm{EdS}$ or $v_\textrm{EdS}$.}  The AAG model can actually
be thought of as a swiss-cheese model of holes of radius $v_\textrm{EdS}$. 
(See Ref.\ \cite{Marra:2007pm} for a discussion of swiss-cheese models.)

In Fig.\ \ref{rbd} we display the functions in term of the comoving coordinate
label $r$ multiplied by $H_0a_0$.  That combination is related to the present
comoving proper distance in units of $H_0^{-1}$,  
\begin{equation}
H_0d_P(\textrm{today}) = H_0\int_0^r\frac{R'(r_1,t_0)}{W(r_1)}dr_1
= H_0a_0\int_0^r\frac{Z(x=1,r_1)}{W(r_1)}dr_1
= \int_0^v\frac{Z(x=1,v_1)}{W(v_1)}dv_1,
\end{equation}
where $v\equiv H_0a_0r$.  The spatial curvature term $W(r)$ is close  to
unity, and $Z$ is of order unity, so to a reasonable approximation
$H_0d_P(\textrm{today})=v$, the EdS result. (The choice $h=0.51$ implies
$H_0^{-1}= 5878$ Mpc.) In Fig.\ \ref{rbd} we show the present proper distance
as a function of $v=H_0a_0r$.

\begin{figure}
\begin{center}
\includegraphics[width=16cm]{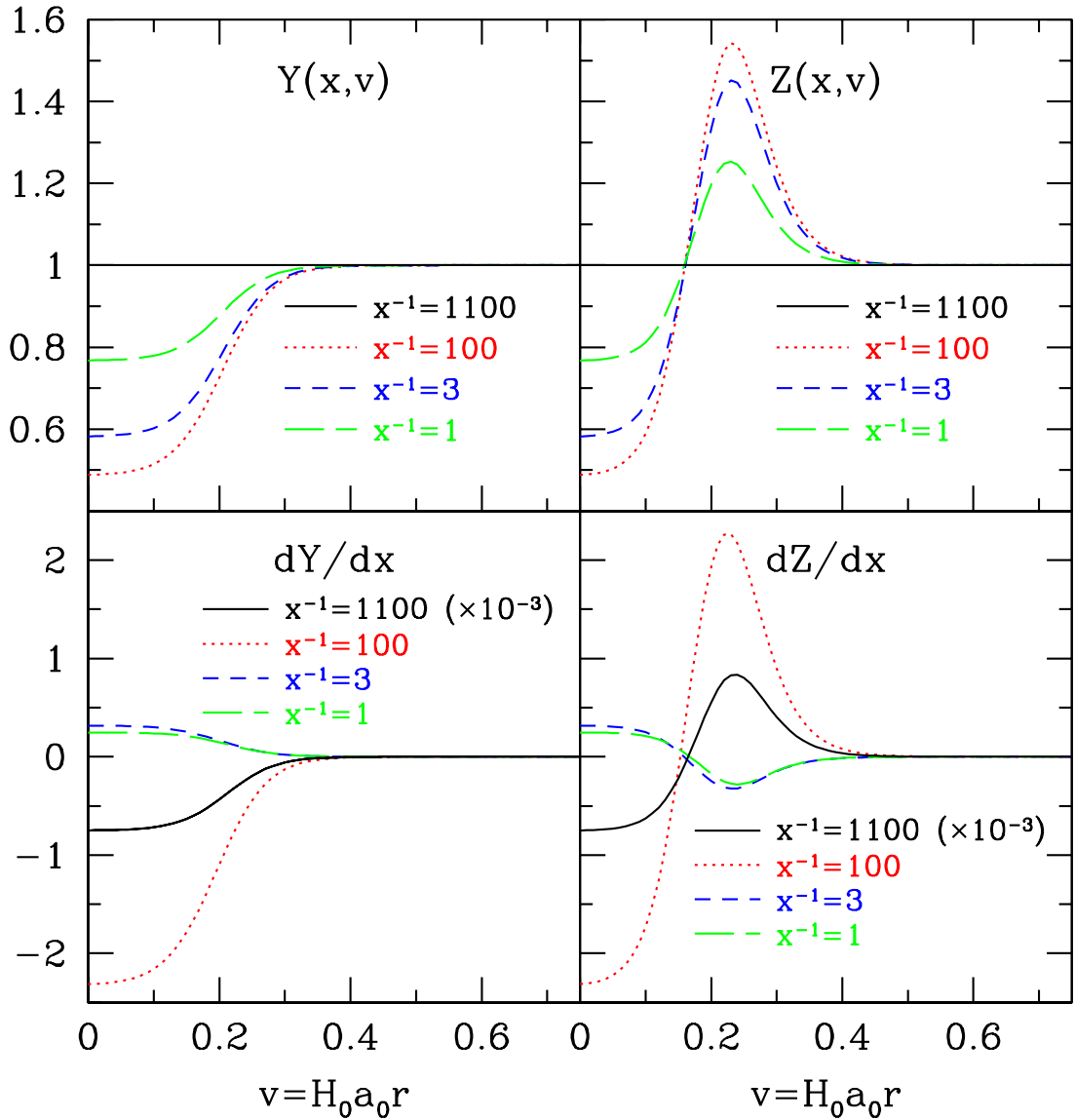}
\caption{Upper figures: The functions $Y$ and $Z$ as a function of $v=H_0a_0r$
for the indicated values of $x=a/a_0$.  At the initial time of the evolution,
$x^{-1}=1100$ and $Y=Z=1$. For the other values of $x$, $Y$ is always smaller than unity
and asymptotically approaches $Y=1$ (the EdS solution) for large values of $v$.
For  small values of $v$ the function $Z$ is smaller than unity, then for larger values
of $v$ it becomes larger than unity, and eventually $Z$  approaches unity from
above. Lower figures: The functions $dY/dx$ and $dZ/dx$ as a function of $v$
for the indicated values of $x=a/a_0$. The function $dY/dx$ never changes sign,
while the function $dZ/dx$ does.  Note that for $x^{-1}=1100$, the values of
$dY/dx$ and $dZ/dx$ have been multiplied by $10^{-3}$.\label{yzypzp} }
\end{center} 
\end{figure}

The evolution of $Y(x)$ and $Z(x)$ is shown in Fig.\ \ref{yzypzp}.  The initial
conditions specified at $x^{-1}=1100$ are $Y=Z=1$, with an initial underdensity
in the interior.  Since the underdense region expands more rapidly than
average, a typical LTB evolution will lead to shell crossing in a rather small
dynamical time.  The occurrence of shell crossing means that the synchronous
coordinate system becomes singular and the evolution in that system can no
longer be calculated.  AGG delay the onset of shell crossing by endowing the
shells with a large initial infall velocity (negative $dY/dx$ and $dZ/dx$). 
Thus, it takes the system a while for the underdense regions with large
expansion rate to reverse the infall and this essentially delays the eventual
shell crossing.\footnote{As far as the purpose of this model is concerned
(having observables similar to the $\Lambda$CDM model), the synchronous gauge
does not have problems, and, in particular, the LTB solution is an exactly
solvable model with nonlinear inhomogeneities for an extended time evolution.}

Alnes, Amarazguioui, and Gr{\o}n also calculate the luminosity distance as a
function of redshift, assuming that the observer is in the center of the
underdense region ($r=0$).  The photon geodesic equation for the position of
the photon as a function of time, $\hat{t}(r)$ or $\hat{x}(r)$, is found from
the LTB metric \cite{Alnes:2005rw}:
\begin{equation}
\label{xofv}
\frac{d\hat{t}}{dr}=-\frac{R'(r,\hat{t})}{\sqrt{1+\beta(r)}}\rightarrow
\frac{d\hat{x}}{dv}=-\hat{x}^{1/2}\frac{Z(\hat{x},v)}{W(v)} ,
\end{equation}
where we have used the new comoving radial coordinate $v=H_0a_0r$.  The 
redshift of the photon, $z(r)$ or $z(v)$, is \cite{Alnes:2005rw,Iguchi:2001sq}
\begin{equation}
\label{zofv}
\frac{dz}{dr}=(1+z)\frac{\dot{R}'(r, \hat{t})}{\sqrt{1+\beta(r)}}
\rightarrow \frac{dz}{dv}=\frac{1+z}{\hat{x}^{1/2}}\frac{Z(\hat{x},v)}{W(v)}
\left(1+\frac{\hat{x}}{Z(\hat{x},v)}\frac{dZ(\hat{x},v)}{d\hat{x}}\right).
\end{equation}
Equations (\ref{xofv}) and (\ref{zofv}) are solved with initial conditions
$z=v=0$ at $x=1$. The luminosity distance is then simply
\begin{equation}
d_L(z)=(1+z)^2R(r,\hat{t}) \rightarrow H_0d_L(z) 
= (1+z)^2 \hat{x}vY(\hat{x},v).
\end{equation} 

\begin{figure}
\begin{center}
\includegraphics[width=16cm]{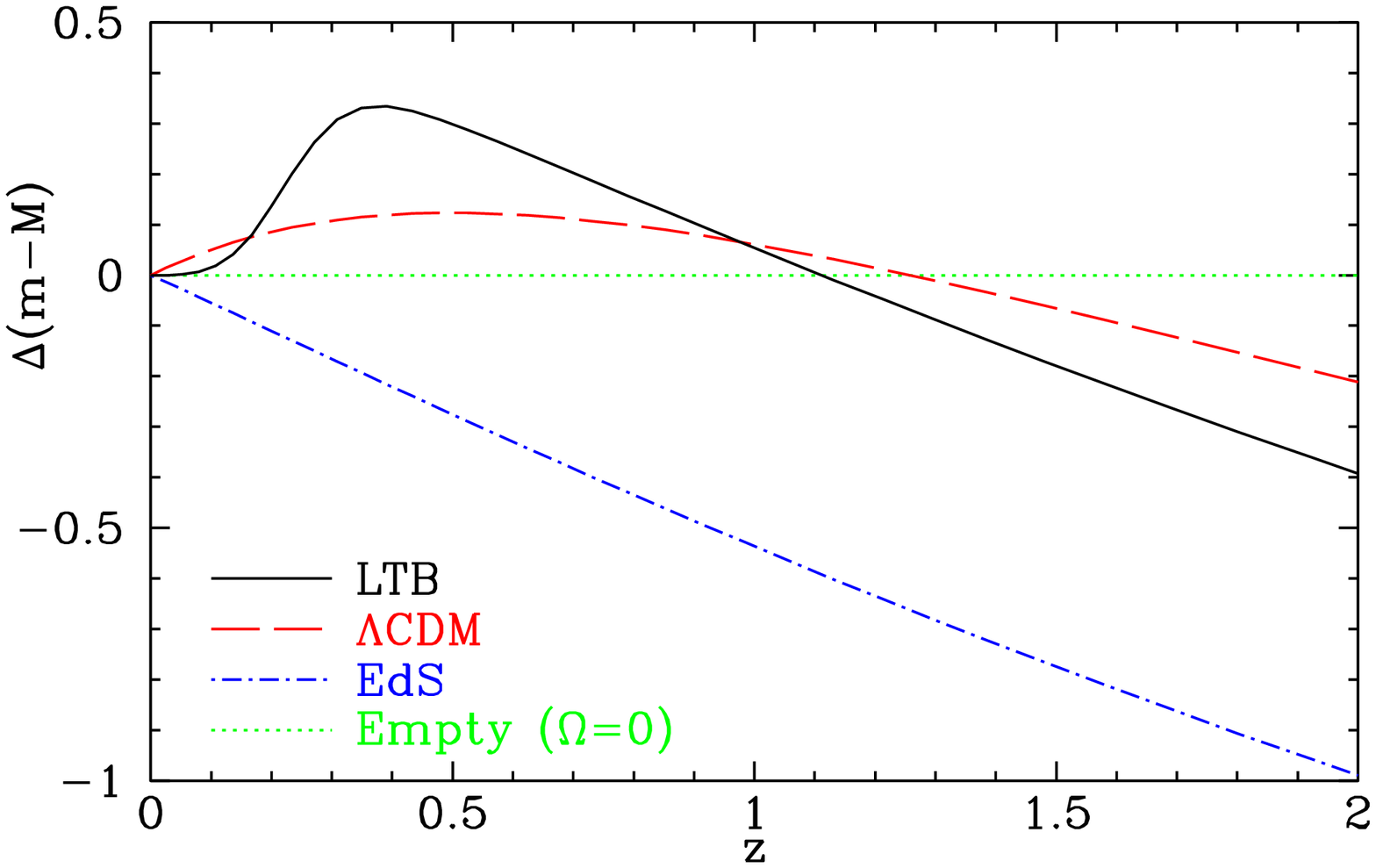}
\caption{The luminosity distance as a function of redshift.  The curve labeled 
LTB is the LTB model of this paper, the model labeled $\Lambda$CDM is the
standard $\Lambda$CDM model with $\Omega_\Lambda=0.7$, the model labeled EdS is
the EdS model, and the model labeled Empty is an empty universe model.  Rather
than $H_0d_L(z)$, we show the usual difference in the distance modulus compared
to the empty model. \label{dlz} }
\end{center}
\end{figure}

A graph of the luminosity distance as a function of redshift is shown in Fig.\
\ref{dlz}.  The striking feature in Fig.\ \ref{dlz} is that the LTB model
resembles a dark-energy model.  AAG show that their LTB model is a better fit
to the ``gold'' data set of Riess \textit{et al.} \cite{Riess:2004nr} than the
concordance $\Lambda$CDM model (at the expense of additional parameters).  With
our different choice for $\Delta r$, $d_L(z)$ in our model is slightly
different than the AGG choice of parameters, but nevertheless demonstrates a
shape characteristic of $\Lambda$CDM models.  Note that as $z\rightarrow0$,
$d_L(z)$ follows the curve expected for the empty model.  This is because the
center of the void \textit{is}, in fact, evolving as an empty universe at late
times.

Whether our (or AAG's) LTB model is a better fit than $\Lambda$CDM in light of
new data is inconsequential. Our goal is not to champion any particular LTB
model as a cosmological model that describes our universe, but rather to
explore the failure of the no-go argument that inhomogeneities can not describe
$d_L(z)$ because the spacetime of our universe can be expressed in the
perturbed Newtonian form.  

We note that in the LTB model there is no dark energy, and since there is only
a pressureless component to the stress-tensor, no individual fluid element
undergoes local acceleration.  

\textit{We emphasize that observations can not directly establish that there
is dark energy, or even that the universe accelerates in the usual sense. 
They only tell us that high-redshift supernovae are about one-half magnitude
fainter than expected in the Einstein-de Sitter model.} Stronger statements
have additional (always unstated) model assumptions!

In the next section we will discuss the formulation of the model using a
perturbed conformal Poisson metric.  But before turning to the main issue of
this paper, we note that although no fluid element in the LTB model undergoes
local acceleration, one can easily show that in the backreaction proposal a
suitably volume-averaged definition of the scale factor has negative pressure
(see Refs.\ \cite{Kolb:2005da, Notari:2005xk, Rasanen:2006kp, Marra:2008sy,
Mattsson:2007tj, Paranjape:2006ww, Wiltshire:2007zj} for a discussion).  This
is related to the averaging issue in cosmology as emphasized by Buchert and
Ellis \cite{Ellis:2005uz, ellis}.

\section{Lema\^{\i}tre--Tolman--Bondi Models as a perturbed conformal Poisson 
metric \label{LTBPG}}

The LTB model of the previous section is an example of a nonlinear cosmological
model.  It is also a model that exhibits an observable that is very different
from the unperturbed model (EdS).  Our goal now is to investigate where and why
the ``no-go'' argument discussed in the Introduction fails.

To get insights, we will express the LTB metric, Eq.\ (\ref{ltbmetric}), in the
form of a  perturbed conformal Poisson metric with small potentials. We will
consider a metric perturbed about the EdS (spatially flat) background. In the
metric we adopt, vector and tensor perturbations are absent: the perturbations
are given by the potentials $\psi$ and $\phi,$\footnote{The perturbed conformal
Poisson metric is closely related to the perturbed conformal Newtonian metric
of Eq.\ (\ref{newtpert}). The differences are that in the perturbed conformal
Poisson metric the spatial perturbation ($\phi$) need not equal the temporal
perturbation ($\psi$), and the perturbed conformal Newtonian metric also has
the additional restrictions  on the metric coefficients discussed after Eq.\
(\ref{newtpert}).}
\begin{equation} 
\label{nmetric}
ds^{2}=a^{2}(\tau) 
\left[-(1+2 \psi) d\tau^{2}+(1-2 \phi)dr^{2}+(1-2 \phi)r^{2}d\Omega^{2}\right],
\end{equation}
where $\tau$ is conformal time, $ad\tau=dt$.

When considering a dust model (or, in general, in the absence of anisotropic
stress), one usually takes the same scalar perturbation in both the spatial and
temporal part of the metric, that is, $\psi=\phi$. Also, it is usually assumed
that the potentials are time independent.

As we will soon show, however, we will have to allow the two potentials to
differ and be time dependent: this is caused directly by the nonlinear
structure of the LTB model. This can be seen mathematically by the fact that in
the case $\psi(r)=\phi(r)$, the system of equations will be overconstrained if
we are beyond linear order. The cause of why $\psi \neq \phi$ and they are time
dependent is actually one of the main issues we will try to understand in this
paper: what we want to show is that it is not possible to describe the AAG
model by means of the perturbed conformal Newtonian metric of Eq.\
(\ref{newtpert}) as far as the linearized theory is concerned.

Of course, this behavior is due to the fact that the AAG model has an initial density and curvature perturbation which is not within the linear regime of perturbation theory.
Even though this is the specific cause of the particular behavior of the AAG model, we will try to learn the general physical cause of that behavior.
The AAG model will be just a tool to understand the general lesson about the importance of voids not dynamically restricted which, as we will explain, are a key-feature to evade the no-go argument discussed in the Introduction.

We point out that the calculations we are going to show are consistent with the
standard result that at linear order the potentials are time-independent and
identical: $\psi(r)=\phi(r)$. It is indeed possible to show (see Ref.\
\cite{Biswas:2007gi}) that if there is no time dependence, then the equations
will not overconstrain the scalar potentials in the case $\psi=\phi$.

In the next subsection we will set up the gauge transformation needed to
calculate the potentials $\psi$ and $\phi$ of the metric of Eq.\
(\ref{nmetric}) in the perturbed conformal Poisson metric corresponding to the
LTB metric of Eq.\ (\ref{ltbmetric}) in the synchronous gauge.

\subsection{Gauge transformation}

To be able to compare the metrics of Eqs.\ (\ref{ltbmetric}) and
(\ref{nmetric}), we will express the latter in the synchronous gauge. We will
use the gauge transformations of the linearized theory: the consistency of this
will be discussed later. 

The generic form of the perturbed EdS model, both in Cartesian and
spherical coordinates, is \cite{Ma:1995ey}
\begin{eqnarray} 
\label{wmetric}
ds^2 & = & a^2(\tau)\left\{-(1+2 \psi)d\tau^2 
- 2 \partial_i \omega \: dx^i d\tau 
+\left[(1-2 \phi)\delta_{i\, j}+D_{i \, j}\chi\right]dx^idx^j\right\} 
\nonumber \\
& = & a^2(\tau) \left[-(1+2 \psi) d\tau^2 - 2\omega'dr d\tau 
+ \left(1-2\phi+\frac{2}{3} \Upsilon \right)dr^2 + 
\left(1-2 \phi-\frac{1}{3}\Upsilon\right)r^2 d\Omega^2\right] ,
\end{eqnarray}
where $D_{i \, j}=\partial_{i}\partial_{j}-\frac{1}{3}\delta_{i \, j}
\nabla^{2}$ and $\Upsilon=\chi''-\chi'/r$. Because of spherical symmetry we
are including only longitudinal degrees of freedom in the metric
perturbations. 

Under a gauge transformation the metric perturbations transform according to
\begin{eqnarray}
\tilde{\psi}&=&\psi-\partial_{\tau} \zeta-\frac{\partial_{\tau}a}{a}\zeta 
\label{psitilde}\\
\tilde{\omega}&=&\omega+\zeta+\partial_{\tau}\beta 
\label{omegatilde}\\
\tilde{\phi}&=&\phi-\frac{1}{3}\nabla^{2}\beta+\frac{\partial_{\tau}a}{a}
\zeta \label{phitilde} \\
\tilde{\chi}&=&\chi+2 \beta, 
\label{chitilde} 
\end{eqnarray}
where $\zeta$ and $\beta$ express the gauge freedom. In radial coordinates,
$\nabla^{2}\beta=\beta''+2 \beta'/r$. This transformation corresponds to a
diffeomorphism which keeps the perturbations in Eq.\ (\ref{wmetric}) small: 
this is actually the definition of gauge transformations in the linearized 
theory.  The new coordinates are given by
\begin{eqnarray}
\tilde{\tau} & = &\tau+\zeta \nonumber \\
\tilde{\vec{x}} & = & \vec{x}- \vec{\nabla} \beta  \nonumber \\
\tilde{r} & = & r-\beta' , \label{trasfco}
\end{eqnarray}
where again only the longitudinal component has been taken into
account.\footnote{Only $\beta'$ will appear in the equations since the 
displacement is $dr=-\beta'$.}

In the following, $\tilde{\psi}, \tilde{\omega},\tilde{\phi},\tilde{\chi},
\tilde{\tau},\tilde{\vec{x}}$, and $\tilde{r}$ will refer to the quantities in
the LTB synchronous gauge, while the corresponding quantities without the tilde
will refer to quantities in the perturbed conformal Poisson metric.

We can now choose the gauge transformations in order to end up with 
$\tilde{\psi}=\tilde{\omega}=0$, that is, to end up in the synchronous gauge.
Taking in account that we are coming from a metric of the form Eq.\
(\ref{nmetric}), which has $\omega=0$ and $\psi\neq0$, we find from Eqs.\
(\ref{psitilde}) and (\ref{omegatilde}) that
\begin{eqnarray}
\tilde{\psi} & = & 0 \rightarrow \psi = \partial_{\tau} \zeta
+\frac{\partial_{\tau}a}{a}\zeta =  (a \, \zeta)\dot{} \\
\tilde{\omega} & = & 0 \rightarrow \zeta = -\partial_{\tau}\beta 
=-a \dot{\beta}.
\end{eqnarray}

These expressions can be expressed in integral form as
\begin{eqnarray}
\zeta(r,t) & = & \frac{1}{a(t)} \int_{\bar{t}}^{t} \psi(r,t_1) \: dt_1
+ \frac{\bar{a}}{a(t)}\bar\zeta(r) \label{zetaz} \label{zetart}\\
\beta(r,t) & = & -  \int_{\bar{t}}^{t}\frac{dt_2}{a^2(t_2)}  
\int_{\bar{t}}^{t_2} \psi(r,t_1) \: dt_1 
- 3 \bar{\zeta}(r) \frac{\bar{t}}{\bar{a}}\left( 1 
- \frac{\bar{t}^{1/3}}{t^{1/3}}\right)
+\bar{\beta}(r) , \label{betart}
\end{eqnarray}
where the overbar denotes the quantity evaluated at $t=\bar{t}$, the initial
time of the evolution ($\bar{t}$ corresponds to $x^{-1}=1100$).

The time-independent integration functions $\bar{\zeta}(r)$ and 
$\bar{\beta}(r)$ express the residual gauge freedom, which we are going to
discuss shortly.  But before doing that, let us make use of the remaining two
gauge transformations, Eqs.\ (\ref{phitilde}) and (\ref{chitilde}).

Comparing the LTB metric, Eq.\ (\ref{ltbmetric}), with the form of the
perturbed metric, Eq.\ (\ref{wmetric}), we see 
\begin{eqnarray} 
1-2\tilde{\phi} + \frac{2}{3}\tilde{\Upsilon} & = & \frac{R'^2}{a^2 W^2}  \\ 
1 -2\tilde{\phi} - \frac{1}{3}\tilde{\Upsilon} & = & \frac{R^2}{a^2 r^2} .
\end{eqnarray} 
Now compare the perturbed conformal Poisson metric, Eq.\ (\ref{nmetric}), with
the form of the perturbed metric, Eq.\ (\ref{wmetric}), with result that
$\Upsilon=\chi''-\chi'/r= 0$. The general solution to this equation is
$\chi=C_1r^2+C_2$, where $C_1$ and $C_2$ are constant in $r$ (but possibly
time-dependent).  In our case we are describing a model where
$\chi\rightarrow0$ as $r\rightarrow\infty$, \textit{i.e.,} it approaches the
unperturbed EdS solution at large $r$, so we must choose $C_1=C_2=\chi=0$. 
Therefore $\Upsilon=0$.

Since we know $\tilde\phi$ and $\tilde\Upsilon$, we can write
\begin{eqnarray}
1 - 2\phi + 2\beta'' - \frac{4}{3}\frac{a}{t}\zeta & = & \frac{R'^2}{a^2 W^2}
\label{btau1}\\
1 - 2\phi +2\frac{\beta'}{r} - \frac{4}{3}\frac{a}{t}\zeta & = & 
\frac{R^2}{a^2 r^2} .
\label{btau2}
\end{eqnarray}
Since we know $\zeta$ and $\beta$ in terms of the potentials and initial data
for $\zeta$ and $\beta$ from Eqs.\ (\ref{zetart}) and  (\ref{betart}), we
finally find the dynamical equations for the potentials $\psi(r,t)$ and
$\phi(r,t)$:
\begin{eqnarray}
1 & - & 2\phi(r,t) - 2\int_{\bar{t}}^{t}\frac{dt_2}{a^2(t_2)}  
\int_{\bar{t}}^{t_2} \psi''(r,t_1) \: dt_1 
+ 6\bar{\zeta}''(r)\frac{\bar{t}}{\bar{a}}
             \left(1 -\frac{\bar{t}^{1/3}}{t^{1/3}}\right) 
+ 2\bar{\beta}''(r) 
- \frac{4}{3t}\int_{\bar{t}}^{t} \psi(r,t_1) \: dt_1
\nonumber \\
& - & \frac{4}{3}\frac{\bar{a}}{t}\bar\zeta(r)  = 
\frac{R'^2(r,t)}{a^2(t) W^2(r)}  \label{dy1}\\
1 & - & 2\phi(r,t) - 2\int_{\bar{t}}^{t}\frac{dt_2}{a^2(t_2)}  
\int_{\bar{t}}^{t_2} \frac{\psi'(r,t_1)}{r} \: dt_1 
+6 \frac{\bar{\zeta}'(r)}{r} \frac{\bar{t}}{\bar{a}}
	\left( 1 - \frac{\bar{t}^{1/3}}{t^{1/3}}\right) 
+ 2 \frac{\bar{\beta}'(r)}{r} 
- \frac{4}{3t}\int_{\bar{t}}^{t} \psi(r,t_1) \: dt_1
\nonumber \\
& - & \frac{4}{3}\frac{\bar{a}}{t}\bar\zeta(r) =
\frac{R^2(r,t)}{r^2a^2(t)} \label{dy2}.
\end{eqnarray}
These are the dynamical equations to solve for $\phi(r,t)$ and $\psi(r,t)$.

Before solving, it is illustrative to evaluate Eqs.\ (\ref{dy1}) and
(\ref{dy2}) at the initial time $t=\bar{t}$ to find the integration functions
$\bar{\zeta}(r)$ and $\bar{\beta}'(r)$.  The result is
\begin{eqnarray}
\bar{\zeta}(r) & = & -\frac{3}{2}\frac{\bar{t}}{\bar{a}}
\left[\bar{\phi}(r)-\int_r^\infty\frac{W^2(r_1)-1}{2r_1W^2(r_1)}dr_1\right] 
\label{zebeze}\\
\bar{\beta}'(r) & = & r \int_r^\infty\frac{W^2(r_1)-1}{2r_1W^2(r_1)}dr_1.
\label{bebebe}
\end{eqnarray}

The fact that $\bar{\zeta}(\infty)=0$ was used in Eq.\ (\ref{zebeze}): all the
gauge quantities have to go to zero as $r\rightarrow\infty$ where the
spacetime is exact Einstein-de Sitter.

The residual gauge freedom expressed by $\bar{\zeta}(r)$ and $\bar{\beta}'(r)$
is therefore fixed by the value of $\phi$ at initial time, that is, by the
initial conditions of the hole in the synchronous gauge.

Note that $\beta'$ must vanish at the center of the hole because of spherical
symmetry.  Only $\zeta$ is allowed to be nonzero at $r=0$.

\clearpage

\subsection{Potentials} \label{potentials}

With some manipulation we can express Eqs.\ (\ref{dy1}) and (\ref{dy2}) in a
more manageable form.  The difference of Eqs.\ (\ref{dy1}) and (\ref{dy2})
yields one equation, while combining Eq.\ (\ref{dy1}) with Eq.\ (\ref{dy2})
multiplied by $r$ and then differentiated with respect to $r$ yields a second
equation:
\begin{eqnarray}
\int_{\bar{t}}^t\frac{dt_1}{a^2(t_1)}\int_{\bar{t}}^{t_1}
\left(\frac{\psi'(r,t_2)}{r}\right)'  dt_2
+3\frac{\bar{t}}{\bar{a}}\left(1-\frac{\bar{t}^{1/3}}{t^{1/3}}\right)
	\left(\frac{\bar{\zeta}'(r)}{r}\right)'
-\left(\frac{\bar{\beta}'(r)}{r}\right)' & = & V(r,t) \label{ni1} \\
\phi'(r,t) + \frac{2}{3}\frac{1}{t}\int_{\bar{t}}^t\psi'(r,t_1) dt_1
+\frac{2}{3}\frac{\bar{a}}{t}\bar{\zeta}'(r) & = & S(r,t), \label{ni11}
\end{eqnarray}
where the functions $V(r,t)$ and $S(r,t)$ are defined as
\begin{eqnarray}
V(r,t) & = &
\frac{1}{2r}\left(\frac{R^2(r,t)}{a^2(t)r^2} -
	\frac{R'^2(r,t)}{a^2(t)W^2(r)}\right) \\
S(r,t) & = & 
\frac{1}{2r}\left(\frac{R^2(r,t)}{a^2(t)r^2} +
	\frac{R'^2(r,t)}{a^2(t)W^2(r)}-
	\frac{2R(r,t)R'(r,t)}{a^2(t)r} \right) . \label{ni2}
\end{eqnarray}	
Using the fact that as $r\rightarrow\infty$, the potentials and their
derivatives must vanish, we can integrate Eqs.\ (\ref{ni1}) and (\ref{ni11})
with respect to $r$ to obtain
\begin{eqnarray}
\int_{\bar{t}}^t\frac{dt_1}{a^2(t_1)}\int_{\bar{t}}^{t_1} \psi(r,t_2) dt_2 
	+3\frac{\bar{t}}{\bar{a}}\left(1-\frac{\bar{t}^{1/3}}{t^{1/3}}\right)
	\bar{\zeta}(r) -\bar{\beta}(r) 
	& = & \int_r^\infty dr_1r_1\int_{r_1}^\infty V(r_2,t)dr_2 \\
\phi+\frac{2}{3}\frac{1}{t}\int_{\bar{t}}^t \psi(r,t_1) dt_1
	+\frac{2}{3}\frac{\bar{a}}{t}\bar{\zeta}(r) & = &
	- \int_r^\infty S(r_1,t)dr_1 .
\end{eqnarray}
Final manipulations involving multiple differentiations with respect to $t$ and
integration by parts yield the results:
\begin{eqnarray}
\psi(r,t) & = & \frac{1}{2}\int_r^\infty (r_1^2-r^2)\dot{Q}(r_1,t) dr_1 
	\label{finalpsi} \\
\phi(r,t) & = & 
     -\frac{1}{2}\frac{\dot{a}}{a}\int_r^\infty (r_1^2-r^2)Q(r_1,t) dr_1
     -\int_r^\infty S(r_1,t)dr_1, \label{finalphi}
\end{eqnarray}
where $Q(r,t)=a^2(t)\dot{V}(r,t)$.

Note that the integration functions $\bar{\zeta}(r)$ and $\bar{\beta}'(r)$ have
disappeared from the final solution. They are, however, determined by the value
of $\phi(r,\bar{t})$ at initial time as seen from Eqs.\  (\ref{zebeze}) and
(\ref{bebebe}). For completeness we give the expressions for $\zeta(r)$
and $\beta'(r)$:
\begin{eqnarray}
\zeta(r,t) & = & \tilde{\tau}-\tau = \frac{1}{2 a} 
\int_r^\infty a^2(r_1^2-r^2) \; \dot{V}(r_1,t) \, dr_1
  \label{zerta} \\
\beta'(r,t) & = & r-\tilde{r} = r \int_r^{\infty} V(r_1,t) \, dr_1 
  \label{berta} .
\end{eqnarray}

Now we are in a position to express the potentials, the displacements in time,
$\Delta t=a\zeta(r,t)$, and radial coordinate, $\Delta r =\beta'(r,t)$, in
terms of the new radial coordinate $v\equiv H_0a_0r$.  In terms of the
variables $Y$ and $Z$ defined in Sec.\ \ref{theltb}, the expressions are
\begin{eqnarray}
2\psi(v,x) & = & \int_v^\infty(v_1^2-v^2)M(x,v_1) \frac{dv_1}{v_1} \\
2\phi(v,x) & = & -\int_v^\infty(v_1^2-v^2)N(x,v_1) \frac{dv_1}{v_1} 
	- \int_v^\infty P(x,v_1) \frac{dv_1}{v_1}   \label{phinew} \\
2\frac{\Delta t}{t} & = & 
	\frac{3}{2}\int_v^\infty(v_1^2-v^2)N(x,v_1)\frac{dv_1}{v_1} \\
2\frac{\Delta r}{r} & = & 
	\int_v^\infty L(x,v_1)\frac{dv_1}{v_1} .
\end{eqnarray}
The functions $M(x,v)$, $N(x,v)$, $P(x,v)$, and $L(x,v)$ are 
\begin{eqnarray}
M(x,v) & = & x\left(\frac{dY(x,v)}{dx}\right)^2 - 
	\frac{x}{W^2(v)}\left(\frac{dZ(x,v)}{dx}\right)^2 
	+ \frac{Z^2(x,v)}{2xW^2(v)}
		\left(\frac{A_2(v)}{Y^2(x,v)Z(x,v)} -2\frac{A_1(v)}{Y^3(x,v)}
			-1\right) \nonumber \\
	& & -\frac{Y^2(x,v)}{2x}\left(\frac{A_1(v)}{Y^3(x,v)}-1\right)
      \\
N(x,v) & = & Y(x,v)\frac{dY(x,v)}{dx}
	-\frac{Z(x,v)}{W^2(v)}\frac{dZ(x,v)}{dx} \\
P(x,v) & = & Y^2(x,v) + \frac{Z^2(x,v)}{W^2(v)}-2Y(x,v)Z(x,v) \\
L(x,v) & = & Y^2(x,v)-\frac{Z^2(x,v)}{W^2(v)} ,
\end{eqnarray}
where, as before, $x=a/a_0$.  The evolution equations for $Y$ and $Z$ are
given in Eqs.\ (\ref{dynamicaleqs}).  The initial conditions are $Y(v,\bar{x})
= Z(v,\bar{x}) = 1$ at $\bar{x}^{-1}=1100$. It is clear that for an EdS
solution ($W(v) = Y(v,x) = Z(v,x) = 1$, $A_1(v)=1$, and $A_2(v)=3$) the sources
$M(x,v)$, $N(x,v)$, $P(x,v)$, and $L(x,v)$ vanish.  Since we will never allow
shell crossing in the evolution, each shell evolves separately.

In the next sections we will present the results for the potentials and
coordinate displacements.

\subsection{Results} \label{results}

\begin{figure}
\begin{center}
\includegraphics[width=16cm]{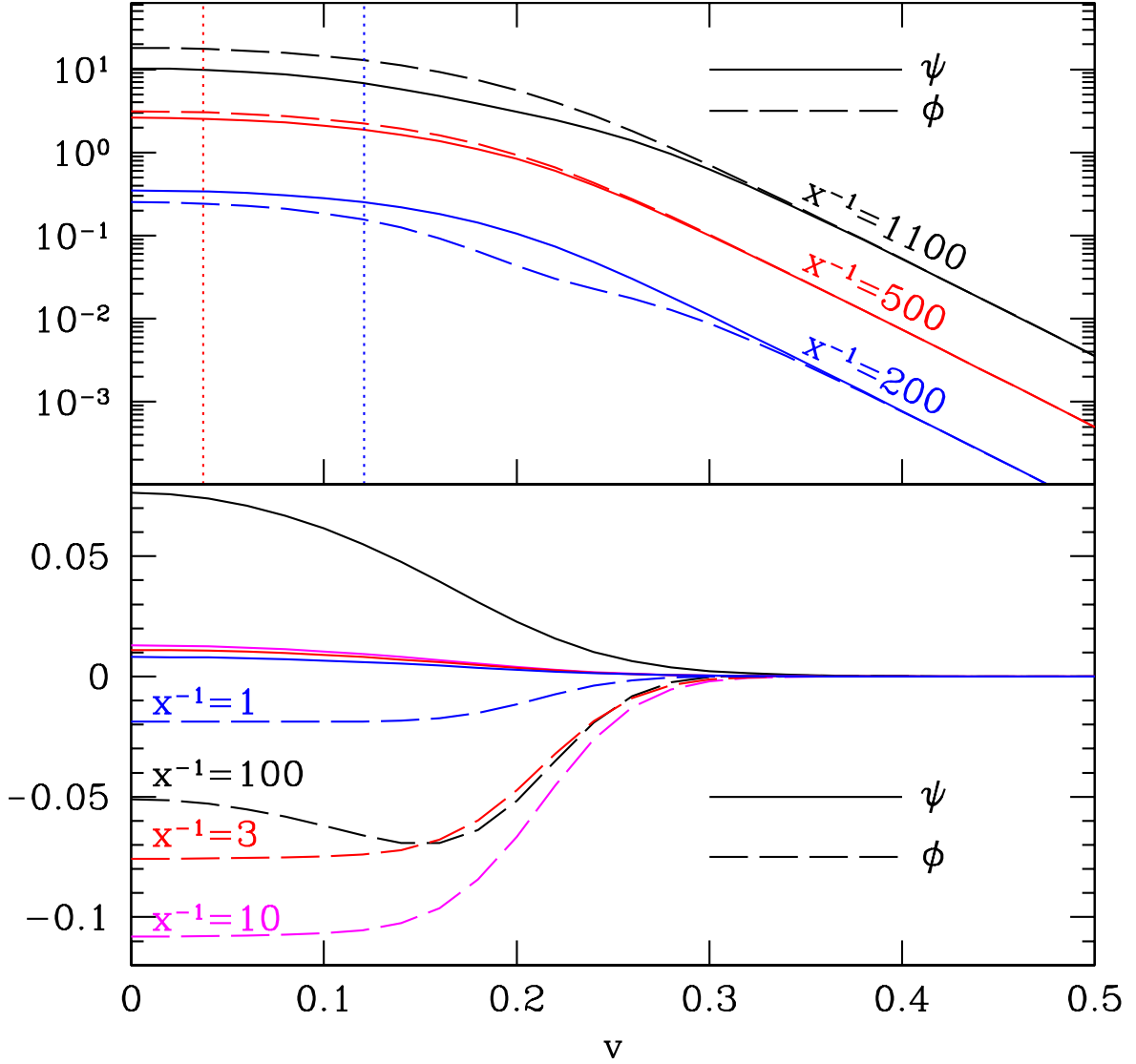}
\caption{The gauge potentials $\psi(x,v)$ and $\phi(x,v)$ as function of $v$
for the indicated values of $x$. The potential $\psi(x,v)$ is indicated by
solid curves and the potential $\phi(x,v)$ is indicated by dashed curves.  In
the lower half of the figure the values of $x$ are indicated only for
$\phi(x,v)$; the values of $x$ for $\psi(x,v)$ are, from top to bottom,
$x^{-1}=100$, $10$, $3$, and $1$. In the upper half of the figure, the values
of $v$ at the horizon are indicated by a vertical dotted line for $x^{-1}=500$
(leftmost line) and $x^{-1}=200$.  \label{pp2} }
\end{center}
\end{figure}

First, we show in Fig.\ \ref{pp2} the values of the potentials $\psi(x,v)$ and
$\phi(x,v)$ for various values of $x=a_0/a$.  There are several striking
features.  First of all, at the initial time, $x^{-1}=1100$ (as well as for
$x^{-1}=500$), $\phi$ and $\psi$ are greater than unity for $v \alt 0.3$.  Also
striking is the fact that near $v=0$, $\phi \neq \psi$.  For $x^{-1}\alt 100$,
$\phi$ and $\psi$ are smaller than unity, but they have different signs.

In the next section we will explain the reason for this behavior.  Here, we
note that when the potentials are greater than unity, they are constant within
the horizon.  If we take the initial epoch to be $x^{-1}=1100$, then $v_H$ is
found by solving Eq.\ (\ref{xofv}).  The value of $v_H$ for the LTB metric is
not too much different than the corresponding value for the EdS model,
$v_H=2x^{1/2}$.  In the EdS model, $v_H=0.06,\ 0.09,\ 0.14,\  0.20,\ 0.6,\
1.15$, and $2$ at $x^{-1}=1100,\ 500,\ 200,\ 100,\ 10,\ 3$, and $1$
respectively. In Fig. \ \ref{pp2} we indicate values of $v_H$ for $x^{-1}=500$
and $200$. As seen in Fig.\ \ref{pp2}, when the potentials are larger than
unity, within the scale of the horizon, they are very nearly constant.  We will
return to this point in Sect.\ \ref{Conclusions}.

Now we turn to Fig.\ \ref{dtdr} where we show the coordinate differences
between the LTB metric and the perturbed conformal Poisson metric.  First, let
us examine $\Delta t$. At the initial point in the evolution $\Delta t/t$ is
much larger than unity and positive.  So even at the very beginning of the
calculation \textit{we can not express the LTB metric in terms of a perturbed
conformal Newtonian metric!} This occurs even though the metric is everywhere
regular, and density perturbations are not highly nonlinear.

\begin{figure}
\begin{center}
\includegraphics[width=16cm]{dtdr.eps}
\caption{The time displacement $\Delta t/t$ and coordinate displacement
$\Delta r/r$ as a function of $v$ for the indicated values of $x$.  The time
displacement $\Delta t$ is negative for  $x^{-1}=1100$, $500$, and $100$.  It
is positive for $x^{-1}=10$, $3$, and $1$. For $x^{-1}=1100$, $\Delta r$ is
positive, for all  other values of $x$, $\Delta r$ is negative. Clearly, 
$\Delta r=r\Delta r/r$ vanishes in the limit $r\rightarrow 0$ as it must.
\label{dtdr} }
\end{center}
\end{figure}

Eventually, when $x$ drops below about $x^{-1}=100$, $\Delta t/t$ becomes
smaller than unity and continues to decrease.  In the next section we will
discuss the cause of this behavior.

Now let us turn to $\Delta r$.  At the initial point in the evolution,
$x^{-1}=1100$, $\Delta r$ is positive.  This means that $r > \tilde{r}$, or the
coordinate label in the perturbed conformal Poisson metric is greater than the
coordinate label in the synchronous gauge. For other values of $x$, $\Delta r$
is negative (\textit{i.e.,} $r < \tilde{r}$, or the coordinate label in the
perturbed conformal Poisson metric is less than the coordinate label in the
synchronous gauge).  

Finally, we turn to the issue of peculiar velocities.  We first remark that the
dust LTB metric of Eq.\ (\ref{ltbmetric}) is comoving: there are \textit{no}
peculiar velocities.  Again denoting a quantity in the synchronous frame by a
tilde, the four-velocity is $\tilde{u}^\mu=(1,\vec{0})$.  However in the
Poisson frame there are peculiar velocities (see also \cite{Ellis:2001ms}).  At linear order, under a
coordinate transformation $\tilde{x}^\mu\rightarrow\tilde{x}^\mu -
\epsilon^\mu$, we have (quantities in the Poisson frame do not carry a
tilde):
\begin{eqnarray}
u^\mu(x) & = & \tilde{u}^\mu(x) -\partial_\nu\epsilon^\mu\tilde{u}^\nu(x)
		+\epsilon^\sigma\partial_\sigma\tilde{u}^\mu(x)\\
u_\mu(x) & = &  \tilde{u}_\mu(x) +\partial_\mu\epsilon^\nu\tilde{u}_\nu(x)
		+\epsilon^\sigma\partial_\sigma\tilde{u}_\mu(x)     .
\end{eqnarray}
From this we find ($\hat{r}$ is a unit vector in the radial direction)
\begin{eqnarray}
u^\mu(x) & = & (1-a\dot{\zeta}-\dot{a}\zeta,\dot{\beta}'\hat{r})   
           =   (1-\psi,\dot{\beta}'\hat{r}) \\
u_\mu(x) & = & (-1-a\dot{\zeta}+\dot{a}\zeta,a^2\dot{\beta}'\hat{r})
           =   (-1-\psi,a^2\dot{\beta}'\hat{r}) . 
\end{eqnarray}
As expected, the change in the four-velocities is of the same order as the
gauge transformation.  At linear order $u^\mu u_\mu=-1$ and
$u_\mu=g_{\mu\nu}u^\nu$.

\begin{figure}
\begin{center}
\includegraphics[width=16cm]{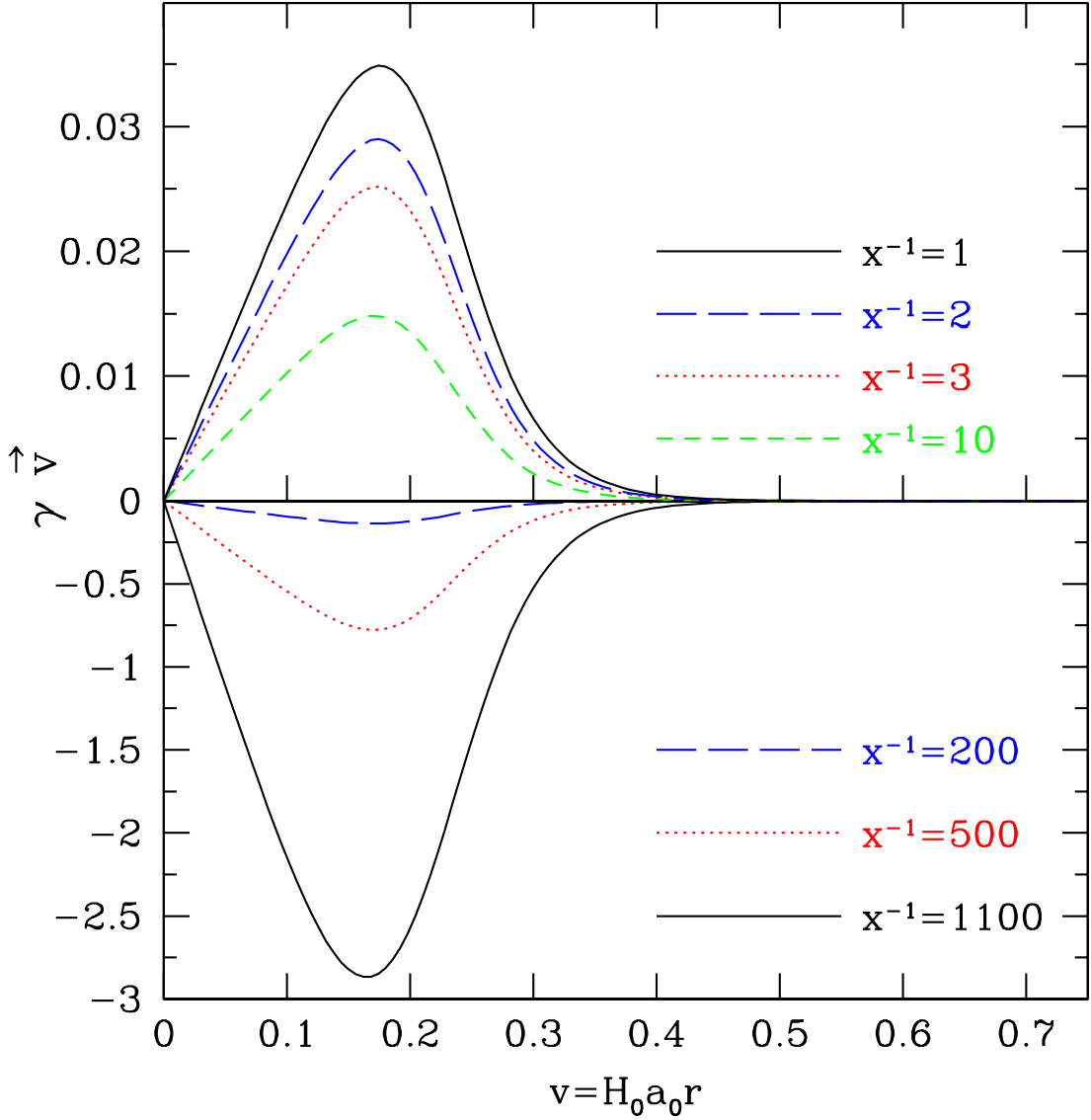}
\caption{The peculiar velocity in the Poisson frame as a function of 
$v=H_0a_0r$ for the indicated values of $x$. A negative (positive) value 
corresponds to a radial velocity towards (away from) the center.  \label{ur} }
\end{center}
\end{figure}

The value of $u^r$ would be interpreted as a peculiar velocity.  Now we express
the four-velocity as $u^\mu=(\gamma,\gamma v^i)$.  The three-velocity is
$\vec{v}=\sqrt{h_{rr}}\,v^r\,\hat{r}=av^r\hat{r}$ at linear order.  Using this,
$\gamma \vec{v}=a\gamma v^r\hat{r}=a\dot{\beta}'\hat{r}$, so
\begin{equation}
\gamma\vec{v}=a\dot{\beta}^\prime \hat{r}= 
		vx^{1/2}\int_{\tilde{v}}^\infty N 
		 \frac{d\tilde{v}}{\tilde{v}} \ \hat{r}.
\end{equation}	
The result for $\gamma \vec{v}$ as a function of $v=H_0a_0r$ is shown in Fig.\ 
\ref{ur} for several values of $x$.  The large initial values of the velocity
are the key to understanding why the perturbative description breaks down.  

We now turn to the discussion.

\section{Discussion\label{Discussion}}

The linear gauge transformation we have performed in the previous Section is a
useful tool to understand where and why the no-go argument discussed in the
Introduction fails.
In this Section, first we will introduce the \textit{general} physical cause of the breakdown of the description by means of the perturbed conformal Poisson metric.
Then we will show how this general feature shows up in the model here examined.
Finally we will conclude the Section with some comments about other models developed in the literature.

\subsection{Lack of a global background}

\begin{figure}
\begin{center}
\includegraphics[width=16cm]{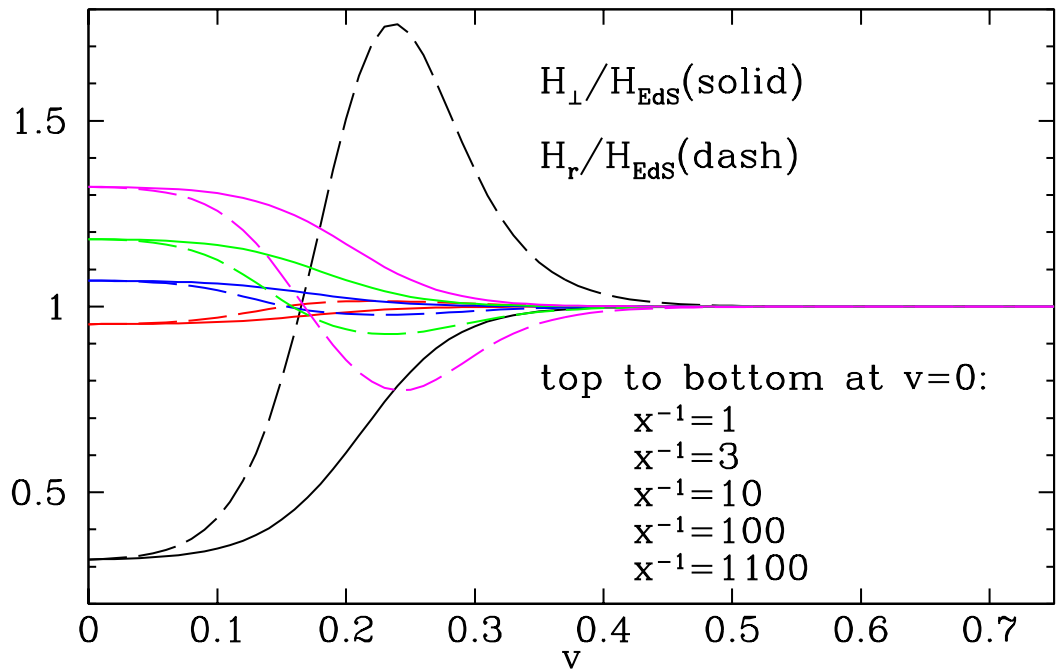}
\caption{The expansion rates as a function of $v$ for the indicated values of
$x$.  \label{h} }
\end{center}
\end{figure}

The description by means of the perturbed conformal Poisson metric breaks down
because of the large $\Delta H$ in the AAG model.  As can be seen from
Fig.\ \ref{h},
\begin{equation}
\frac{\Delta H_{\perp, r}}{H_\textrm{EdS}}= \frac{H_{\perp, r} -H_\textrm{EdS}} 
{H_\textrm{EdS}} 
\end{equation}
is of order unity. A large $\Delta H$ is a typical feature of models with
inhomogeneities, and from this point of view we think the LTB model is a
useful tool to study the consequences of the evolution of structures in the
real universe.  In particular, the AAG version allows us to follow the
evolution of the inhomogeneities for a rather large time span, thanks to the
large initial infall velocity of the shells (negative value of $\Delta H /
H_\textrm{EdS}$ for $x^{-1}=1100$, as can be seen from Fig.\ \ref{h}).

A different expansion rate means a different FLRW solution, and even if the
initial $\Delta H$ is small, the two solutions (void and EdS) will eventually
follow very different evolutions. What we have in mind is the comparison of a
background FLRW solution in the underdensity (the void region) and in the FLRW
solution at large $v$.  We will denote the value of $v$ beyond which the
solution is approximately EdS as $v_\textrm{EdS}$.  We make the point that
these two patches have different backgrounds, even though the ``average''
background \textit{is} the EdS solution. The AAG model can indeed be thought of
as an exact Swiss-cheese model matched to the density and the (zero) curvature
of the EdS solution at a sufficiently large radius ($v \gtrsim
v_\textrm{EdS}$). We can imagine populating the universe with other holes of
radius $v \sim v_\textrm{EdS}$. On scales larger than any spherical hole, the
AAG model evolves as the EdS model. From this perspective, in the framework
developed by Buchert \cite{Buchert:2007ik}, the backreaction of the
inhomogeneities is zero. (See Refs.\ \cite{Marra:2007pm, Marra:2007gc,
Marra:2008xi} for a discussion.)  Here, we find a lack of a global background
in spite of the absence of a volume backreaction.

It can happen that at some specific time the background of the void patch is
close to the EdS solution. This will be, however, a special moment because of
the different evolution of the two patches. It will be more typical to have a
large departure from the EdS flow, which will be manifest in a large $\Delta H$
or in large ``peculiar'' velocities. These ``peculiar'' velocities are the
result of comparing the cosmic evolution of two different FLRW solutions, or,
in other words, the results of perturbing about the wrong background.

Summarizing, \textit{an inhomogeneous universe featuring voids cannot be
studied by means of a  global homogeneous background, and this holds even
if the inhomogeneities do not affect, as it happens in a swiss-cheese model,
the global evolution}.

\subsection{Discussion of the results}

What we have just discussed is clearly shown by the results obtained in the
previous section. To understand them it is useful to focus on two key
quantities. 

First, it is possible to show (see the Appendix \ref{quadro}) that the value of
$\phi$ at the origin has the following time dependence:
\begin{equation}
\label{approx}
\phi_0 \propto - \left( \frac{a_v(t)}{a(t)} -1\right)^2 ,
\end{equation}
where $a_{v}\equiv r^{-1}\left.R \right|_{r=0}$ describes the scale factor of
the void region. In turn, this implies that $\dot{\phi}_0 \propto \Delta
H_\perp$.  This expression makes clear that $\phi_0$ can be thought as the
time integral of $\Delta H$.

A similar relation holds for $\beta'$ (see the Appendix \ref{quadro}): $a \,
\beta' \simeq R - a \, r$.  This implies that $ a \, u^r = a \, \dot{\beta}'
\simeq R \, \Delta H_{\perp} $, where we note that $a \, u^{r}$, the velocity
of the comoving matter in the Poisson frame shown in Fig.\ \ref{ur}, is, as
expected, almost the same  as the ``peculiar'' velocity of the comoving matter
with respect to the EdS background:
\begin{equation} 
\label{pecp}
v_{pec} = \int_0^r dr_1 \frac{R'}{W} \left(H_{r}-H_\textrm{EdS} \right)
\simeq  R \, \Delta H_{\perp} = R \, H_\textrm{EdS}
\frac{\Delta H_{\perp}}{H_\textrm{EdS}} \sim R \, H_\textrm{EdS},
\end{equation}
where the last relation holds because, as we have seen, $\Delta H/H$ is of 
order unity. (We have also used the fact that $W(r)$ is very close to unity.)

Expressing these key-quantities as a function of $\Delta H$, which is the
source of the effects, will help us in understanding the results of the
previous section.

First, at early times ($x^{-1} \lesssim 500$) ``peculiar'' velocities are
larger than $c$, as can be seen in Fig.\ \ref{ur}. The reason can be understood
from Eq.\ (\ref{pecp}): at early times $H_\textrm{EdS}$ is indeed large.  This
highlights the usefulness of the AAG model, which allows us to encompass both
early times with a large expansion rate and the present time.  This also
illustrates the interesting fact that even with smaller $\Delta H$ we could
have large peculiar velocities at early times when one might expect the system
to be accurately described as a perturbation of an EdS universe.  These large
velocities cause the potentials to be greater than unity and the perturbed
conformal Poisson metric to be unable to describe the metric in the void
region.  In the linear gauge transformation this shows up as the fact that
peculiar velocities are of the same order as the gauge transformation. This is
a general breakdown which affects the perturbed conformal Poisson metric even
if we perform a nonlinear gauge transformation \cite{VanAcoleyen:2008cy}.

\begin{figure}
\begin{center}
\includegraphics[width=16cm]{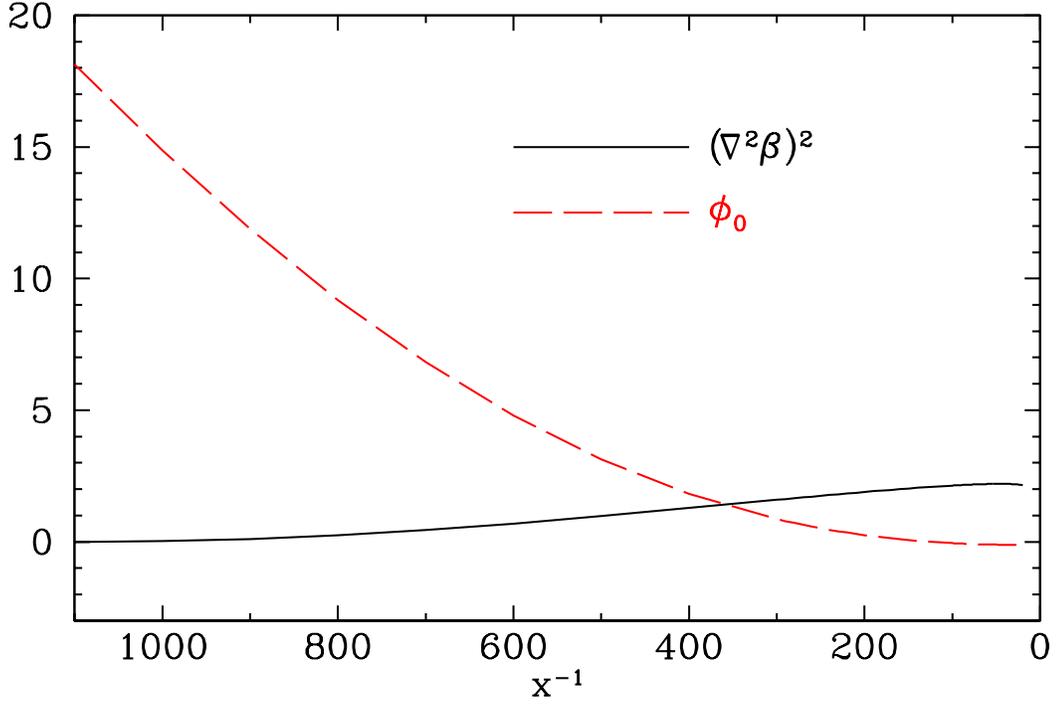}
\caption{The functions $(\nabla^2\beta)^2$ evaluated at $v=0$ (dashed curve) 
and $\phi_0=\phi(v=0,x)$ (solid curve) as a function of $x^{-1}=a_0/a$. 
\label{nb} }
\end{center}
\end{figure}

At later times, velocities become order $10^{-2} c$ as can be seen from Fig.\
\ref{ur}; this is caused by the slowing down of the expansion.  However, both
$\beta'= \Delta r$ and $\phi$ do not depend on $\Delta H$, but rather on its
time integral, or, in other words, on the different size of the void region
compared to the EdS region, that is, on $R- a \, r$. This dependence appears in
$\phi$ because it is the spatial perturbation that has to connect these two
differently expanding spatial regions.  The reason for the breakdown of the
linear transformation is, therefore, the growth of $\beta'= \Delta r$, that is,
the departure of the two metrics.  This shows up directly in $\phi$.  The basic
problem is that the linear gauge transformation neglects terms like ($\nabla^2
\beta)^2$ compared to terms like $\phi$: see Eq.\ (\ref{phitilde}).  As you can
see from Fig.\ \ref{nb}, this approximation is no longer valid for $x^{-1} \alt
400$.  So even if the velocities are small, the linear transformation breaks
down because their time-integrated effect is large. As a consequence we find,
for example, $\phi \neq \psi$. It is however possible to describe the late evolution of the model
in the perturbed conformal Newtonian metric by means of a nonlinear gauge transformation
whose potential follows nonlinear equations of motion \cite{VanAcoleyen:2008cy}. 

\subsection{Comments}

We will comment now on the models of Refs.\ \cite{Paranjape:2008ai} and
\cite{Alexander:2007xx}. According to Ref.\ \cite{Paranjape:2008ai} there is no
breakdown of the perturbed conformal Newtonian metric during the dynamical
evolution of the \textit{particular} (it is important to not generalize) LTB
model the authors have considered. The model of Ref.\ \cite{Alexander:2007xx}
does not deal with the possibility of using the perturbed conformal Newtonian
metric to describe the spacetime of the universe, but it will nevertheless
help us to make our point because it is a model which can be described with the
perturbed conformal Newtonian metric if we perform a nonlinear gauge
transformation.

We start noting that in both Ref.\ \cite{Paranjape:2008ai} and Ref.
\cite{Alexander:2007xx} the validity of the perturbed conformal Newtonian
metric is assumed as a starting point. These particular models are not allowed to
have large peculiar velocities. The point is that this is a particular choice,
and, in our opinion, not general as demonstrated by the dependence in Eq.\
(\ref{pecp}) of the  peculiar velocities on $\Delta H$. We would like to stress
again that these ``background'' peculiar velocities have nothing to do with
anything that can be measured as a local effect. Therefore, the models of
Refs.\ \cite{Paranjape:2008ai} and  \cite{Alexander:2007xx} over-restrict the
dynamics of the inhomogeneities, or, equivalently, demand the existence of a
global background. For example, in the model of Ref.\ \cite{Paranjape:2008ai},
the initial $\Delta H$ of Eq.\ (\ref{pecp}) is exactly zero, and in the model
of Ref.\ \cite{Alexander:2007xx} it is $\Delta H / H_{EdS} < 10^{-3}$.
Therefore, these models  serve as an illustration to show the circular
reasoning at the basis of the no-go argument: if we demand that the spacetime
metric of the inhomogeneous universe has a global background without any
departure due to differently evolving regions, then we will find the validity
of the perturbed conformal Newtonian metric. This is the main point of this
work, and the linear gauge transformation is a useful tool to understand it.

It is interesting to note that the model of Ref.\ \cite{Alexander:2007xx} can
confront the SNe data, but its luminosity distance does not show a big
departure in shape from the EdS one, while the AAG model features a luminosity
distance similar to the $\Lambda$CDM model. This is mainly due to the size of
the hole, roughly $10$ times smaller than the AAG one. We point out that if we
want to have a $\Lambda$CDM-like luminosity distance in a setup with only one
hole and the observer at the center, then we will likely have large background
peculiar velocities. This might, however, be a typical, but not a general
feature. To illustrate this we have calculated the luminosity distance for the
model of Ref.\ \cite{Alexander:2007xx}, but with a hole roughly $10$ times
larger: we found that $d\Delta(m-M)/dz <0$ always.
This can be seen, again, as due to a restriction on the dynamics: allowing larger
peculiar velocities we find voids with more general dynamics. We stress again
the usefulness of the AAG model in showing these general features.

\section{Conclusions \label{Conclusions}}

The LTB model analyzed in this paper is a very useful toy model to explore the
role of large inhomogeneities in the determination of cosmological observables
such as the luminosity distance as a function of redshift. We employed the toy
model to demonstrate where and why the no-go argument discussed in the
Introduction fails. To this end we have performed a linear gauge transformation
in order to understand the problems we have to face if we want to express the
LTB model in the form of a perturbed conformal Newtonian metric, Eq.\
(\ref{newtpert}).

We would like to stress that we regard the AAG model as a tool useful to show
a general feature needed to evade the no-go argument: we focused on the \textit{general}
physical cause of the nonlinear behavior of the AAG model more than on its particular
initial conditions.

We came to the conclusion that the way the no-go argument is proved is by
assuming, as a starting point, the validity of the argument itself. The way to
escape the no-go argument is by allowing large ``background peculiar
velocities,'' where these velocities have nothing to do with, \textit{e.g.,}
velocity dispersions in clusters or anything that can be measured as a local
effect. If indeed inhomogeneities are responsible for the ``apparent''
acceleration of the universe, then the evolution of the expansion rate must
resemble a $\Lambda$CDM model.  If that is true, then there will naturally be
``peculiar'' velocities if one describes the evolution in terms of
perturbations of an EdS model.  One cannot simply assume peculiar velocities
are small and then use that as a basis to show inhomogeneities can have no
effect. In the calculation of peculiar velocities it is necessary to specify a
background solution about which to calculate velocities.  The peculiar
velocities can be small in the $\Lambda$CDM background but not in the EdS
background. The velocities appear as a departure from EdS of the Hubble flow,
preventing the metric from being written in the conformal Newtonian form.

In other words, to evade the no-go argument we need to free the inhomogeneous
models of dynamical restrictions (small background peculiar velocities) which
impose the existence of a global background.

Another way to see how the no-go argument of Eq.\ (\ref{newtpert}) fails is
through the equations of motion: in order to have a model, \textit{e.g.,} an
LTB model, explain the observed departure of $d_L(z)$ from the EdS result
requires nonlinear equations of motion, and this is because of the large
background peculiar velocities which can generally occur. By this, we mean that
in order to end up with a large spatial variation of the background $H$
requires a nonlinear evolution like the one of the LTB model discussed in this
paper. And this can occur even if it is possible, as in the AAG model, to
describe an inhomogeneous universe by means of the perturbed conformal
Newtonian metric from (let's say) $z=2$ to present time.  This is shown here by
explicit calculation.  

Therefore, one must reconsider very carefully the statement that
inhomogeneities can not have a significant effect on cosmological observables
such as the luminosity distance as a function of redshift if arguments are
based on the \textit{assumption} that the metric describing our universe can be
written in the perturbed conformal Newtonian form of Eq.~(\ref{newtpert}).

Now we enumerate some of our findings.

\begin{enumerate}

\item  In the model we consider, even at late times when the density contrast
is linear and the velocities are small, one can not express the LTB metric in
the perturbed conformal Newtonian form of Eq.\ (\ref{newtpert}) by means of a
linear gauge transformation. The breakdown of the linear transformation clearly
shows the importance of background velocities.

The linear gauge transformation breaks down mainly because of the large 
$\Delta r = \beta'$.   This is caused by the different expansion rates ($\Delta
H$) of the void and the EdS solution, as can be seen in Fig.\ \ref{h}, which
shows the expansion rates as a function of $v$ for several values of $x$. The
result is that the two metrics do not remain close enough for a linear
transformation to connect them.  As discussed in the previous section, the
basic problem is that the linear gauge transformation neglects terms like
($\nabla^2 \beta)^2$ compared to terms like $\phi$. Summarizing, the cause of
the break down of the linear gauge transformation is the large $\Delta H$, and
the underlying cause of that is the nonlinearity of the equations of motion.
The failure described above can be stated as the fact that the linearized
equations of motion work only for a limited range of times.

\item  It is possible to find a nonlinear gauge transformation to overcome this
problem with a perturbed conformal Newtonian metric description of the LTB
metric whose potential follows nonlinear equations \cite{VanAcoleyen:2008cy}. These nonlinear
equations (which are actually the exact Einstein equations) at some time will
typically give large potentials and make the perturbed conformal Newtonian
gauge description inconsistent (large $\Delta t = a\zeta$ in the LTB model
examined) because at some point the peculiar velocities will be larger than $c$
if we do not restrict the dynamics, as shown by the AAG model.

\item Although at the present time, in the model we consider, the magnitude of
the potentials are small, they are not equal at the level of the linearized
theory, \textit{viz.}\ $\psi \neq \phi$.  This could na\"{\i}vely be
interpreted as some anisotropic stress or something else funny with the stress
tensor.  This is exactly the point that has been made in, \textit{e.g.,} Ref.\
\cite{Kolb:2005da}: Analyzing an inhomogeneous model in the framework of a
homogeneous model introduces spurious components and behavior to the stress
tensor (\textit{possibly even a negative-pressure fluid!}). 

\item If we start with a perturbed EdS model, then the growth of perturbations
will lead to the formation of large voids, filamentary structure, and other
nonlinearities.  Perhaps the LTB models are useful toy models to study the
effect of that phenomenon.  The model explored in this paper is a good example
because it is capable of dealing with nonlinearities within a large timespan.

\item Finally, another point worth mentioning is the physical nature of
super-Hubble modes of $\phi$ and $\psi$.  At the initial time, within each
Hubble radius patch, $\phi$ and $\psi$ are very nearly constant, but possibly
larger than one (with a value of about $20$ near the origin).  That
\textit{does not} mean that we can simply subtract different values, because we
are only allowed to do this once.  If we do it at $x^{-1}=1100$, then at
$x^{-1}=1$ we would have $\phi$ and $\psi$ near the origin of $-20$.

\end{enumerate}

Summarizing, inhomogeneities, in particular voids not dynamically restricted,
undermine the existence of a global background.

The interesting question for future consideration is if backreactions\footnote{We do not really distinguish between weak or strong backreaction because we think that what matters are the observables. Assuming, indeed, that an inhomogeneous model gives a $\Lambda$CDM-like luminosity distance, then it is not straightforward to understand which category of backreaction (weak or strong) the model belongs to. The strong backreaction indeed is about a change of background, but this is also the case for the weak backreaction because, as in the case of this example, it ``defines'' an effective background, the $\Lambda$CDM model. Moreover, the change of background of the strong backreaction should be related to observable data on the light cone as it is for the weak backreaction.} of cosmological perturbations can substantially distort derived
conclusions about the energy-momentum tensor of the universe without allowing
dramatic violations of \textit{statistical} homogeneity. The present work
addresses the key issue that backreaction effects cannot be assumed \textit{in
principle} to be small on the basis of the no-go argument explained in the
Introduction and therefore opens a wide range of possibilities to tackle the
study of the inhomogeneities in the universe.

\begin{acknowledgments}
It is a pleasure to thank J. M. Bardeen, T. Buchert, S. M. Carroll, K. Van Acoleyen and H. Van Elst for useful discussions, comments and suggestions. This work was supported in part by the
Department of Energy. V.M.\ would like to acknowledge the Kavli Institute for
Cosmological Physics for hosting a visit to the University of Chicago and
``Fondazione Angelo Della Riccia'' for support. S.M.\ and V.M.\ acknowledge ASI
contract I/016/07/0 ``COFIS" for partial financial support. 
\end{acknowledgments}


\appendix*
\section{Details of the approximations used in Sect.\ \ref{Discussion}} 
\label{quadro}

In this Appendix we will discuss the approximations used in Sect.\
\ref{Discussion}. Although we have checked numerically that they hold, we will
present here an analytic explanation.

\subsection{Approximations for $\phi_0$}

We first want to motivate that the time dependence of $\phi$ near the origin
can be estimated as $\phi_0 \sim -(a_v/a-1)^2$, where $a_{v}= r^{-1}\left.R
\right|_{r=0}$, as used in Eq.\ (\ref{approx}).  

The second term dominates the expression for $\phi(r,t)$ in Eqs.\
(\ref{finalphi}) or (\ref{phinew}), so in the limit that $W(r)=1$, $\phi_0(t)$
can be written as
\begin{equation}
\phi_0(t) \simeq - \int_0^{r_{\textrm{EdS}}} \frac{1}{2}
\left(\frac{R}{a \, r_1} - \frac{R'}{a}\right)^{2} \frac{dr_1}{r_1} =
-\int_0^{r_{\textrm{EdS}}} \left[ \left(\frac{R}{a \, r_1} - 1 \right)' 
\right]^{2} \frac{r_1}{2} dr_1 ,
\end{equation}
where  we defined $r_{\textrm{EdS}}$ as the coordinate radius at which the 
metric is the EdS one (it will disappear from the final expression).  To
proceed further, define a function $c(r,t)$ as
\begin{equation}
\phi_0(t) = - \int_0^{r_{\textrm{EdS}}} \left[ \left(\frac{R}{a \, r_1}   
-1 \right)' \right]^{2} \frac{r_1}{2} dr_1 
= - c(0,t) \left [  \int_0^{r_{\textrm{EdS}}}  \left(\frac{R}{a \, r_1} -1 
\right)'  \frac{r_1}{2} dr_1 \right ]^2
\left[ \int_0^{r_{\textrm{EdS}}} \frac{r_1}{2} dr_1 \right]^{-1} .
\end{equation}
Now define $\left.R \right|_{r=0} = a_{v} \, r$ where $a_{v}$ gives the scale
factor for the FLRW solution at the center. Thanks to Birkhoff's theorem, outer
shells do not influence the center, which can be indeed thought as a
homogeneous patch.  Expanding $R(r,t)$ in the void region as\footnote{In the
expansion $a_{v1}=0$, thanks to the fact that $\rho'(0,t)=0$ as discussed in
Ref.\ \cite{Vanderveld:2006rb}.}
\begin{eqnarray}
\frac{R(r,t)}{r} = a_v(t) + a_{v2}(t) \, r^2 + {\cal O}(r^3),
\end{eqnarray}
we find 
\begin{equation}
\phi_0(t) \sim - \left ( \frac{a_{v}}{a} -1\right )^2 ,
\end{equation}
where, after integrating by parts, we have made the approximation that the void
region occupies almost all  of the volume of the hole of radius
$r_{\textrm{EdS}}$.

Now we wish to show that $c(r,t)$ depends only weakly on time. To this end, we
expand $\left( {R / ar} \right)' $ and $\left[ \left( {R / ar} \right)'
\right]^{2}$ in $r$ around $r_{\textrm{EdS}}$ to third order, that is up to the
fourth derivative of $R$ with respect to $r$, and then we plug the resulting
series in the integrals. The result is
\begin{equation}
c(r,t) \simeq \frac{4}{3} + \frac{1}{9}\left(-\frac{4}{r_{\textrm{EdS}}} +
\frac{R'''(r_{\textrm{EdS}},t)}{R''(r_{\textrm{EdS}},t)} \right) 
(r-r_{\textrm{EdS}}) .
\end{equation}
The first remark is that the term proportional to $R''''(r_\textrm{EdS},t)$ 
cancels. The dependence on time is therefore in the $R'''(r_\textrm{EdS},t) /
R''(r_\textrm{EdS},t)$ term. Following a procedure similar to this one used to
obtain Eq.\ (3.5) of Ref.\ \cite{Biswas:2006ub}, but keeping both density and
curvature inhomogeneities, we find:
\begin{equation}
\frac{R'''(r_{\textrm{EdS}},t)}{ R''(r_{\textrm{EdS}},t)}=
\frac{3}{r_{\textrm{EdS}}} + 
\frac{d'''(r_{\textrm{EdS}})}{d''(r_{\textrm{EdS}})},
\end{equation}
where $d(r)=\bar{\rho}(r,\bar{t})+ \bar{d} \, \rho(t)^{-1/3} k(r)$, $\bar{d}$
is a constant, $\rho(t)$ is the EdS density, and $\bar{\rho}(r,\bar{t})=
3\alpha(r)/ 8 \pi R^3(r,\bar{t})$ is the initial averaged density. We see,
therefore, that a general condition to have the constancy of
$R'''(r_{\textrm{EdS}},t) / R''(r_{\textrm{EdS}},t)$ is that the second and
third derivatives with respect to $r$ of $\bar{\rho}(r,\bar{t})$ and $k(r)$ are
proportional, so that the time dependence cancels. This is indeed the case in
the AAG model. We remark that $\bar{\rho}(r,\bar{t})$ and $k(r)$ can be seen as
the defining functions for a LTB model, and this result holds thanks to the
matching guaranteed by $d'(r_{\textrm{EdS}})=0$.

\subsection{Approximations for $\beta'$}

We now discuss the approximation that $a\beta'\simeq R-ar$.  Starting from
Eq.\ (\ref{berta}), with the approximation that $W\simeq 1$, we have
\begin{equation}
\frac{\beta'(r,t)}{r} \simeq \int_r^\infty \frac{dr_1}{r_1} 
\left(\frac{R}{a \, r_1} - \frac{R'}{a}\right) 
\frac{1}{2}\left(\frac{R}{a \, r_1} + \frac{R'}{a}\right)
= -\int_r^\infty dr_1 \left(\frac{R}{ar_1}\right)'
\left[\frac{1}{2}\left(\frac{R}{a \, r_1} + \frac{R'}{a}\right)\right].
\end{equation}
Now from Fig.\ \ref{yzypzp}, the term in the square brackets oscillates about
one, so it is not a bad approximation to set it to unity.  The integral can
then be done with the result that $a \, \beta' \simeq R - a \, r$, used in
Sect.\ \ref{Discussion}.



\end{document}